\documentclass[a4paper,fleqn,usenatbib]{rrl}

\usepackage{newtxtext,newtxmath}

\usepackage[T1]{fontenc}
\usepackage{ae,aecompl}

\usepackage{graphicx}	
\usepackage{amsmath}	
\usepackage{amssymb}	

\title[Extra-tidal RR Lyrae stars in Galactic globular clusters]{Search for Extra-Tidal RR Lyrae Stars in Milky Way Globular Clusters From Gaia DR2}

\author[Kundu et al.]{
Richa Kundu,$^{1}$\thanks{E-mail: richakundu92@gmail.com}
Dante Minniti,$^{2,3,4}$
Harinder P. Singh$^{1}$
\\
\\
$^{1}$Department of physics and astrophysics, University of Delhi, Delhi-110007, India.\\
$^{2}$Instituto Milenio de Astrofisica, Santiago, Chile.\\
$^{3}$Departamento de Ciencias Fisicas, Facultad de Ciencias Exactas, Universidad Andres Bello, Av. Fernandez Concha 700, Las Condes,
Santiago, Chile.\\
$^{4}$Vatican Observatory, V00120 Vatican City State, Italy.
}

\date{Accepted 2018 November 27. Received 2018 November 27; in original form 2018 October 15}

\pubyear{2018}

\begin{document}
\label{firstpage}
\pagerange{\pageref{firstpage}--\pageref{lastpage}}
\maketitle

\begin{abstract}

We used extra-tidal RR Lyrae stars to study the dynamics of Galactic globular clusters and know how effects like dynamical friction and tidal disruption affect these clusters. The Gaia DR2 catalog for RR Lyrae stars \citep{gaiarrl} is used along with the proper motions and tidal radii data for the globular clusters compiled from literature. A sample of 56 Galactic globular clusters is analyzed. Out of these 56 Galactic globular clusters, only 11 have extra-tidal RR Lyrae stars. However, only two clusters, namely, NGC 3201 and NGC 5024, have enough extra-tidal RR Lyrae stars to draw interesting conclusions. NGC 3201 has 13 extra-tidal RR Lyrae stars which are asymmetrically distributed around its center with more number of stars in its trailing zone than its leading part. We conclude that these asymmetrical tidal tails are due to the combined effect of tidal disruption and the stripped debris from the cluster. On the other hand, NGC 5024 has 5 extra-tidal RR Lyrae stars, four of them are concentrated in a region which is at a distance of about 3 times the tidal radius from its center. These may be the stars that are being ripped apart from the cluster due to tidal disruption. The presence of these extra-tidal RR Lyrae stars in the clusters can be an indication that more cluster stars are present outside their tidal radii which may be revealed by deep wide field CMDs of the clusters.

\end{abstract}

\begin{keywords}
(Galaxy:) globular clusters: individual: NGC 3201, NGC 5024 -- stars: variables: RR Lyrae -- Proper motion
\end{keywords}



\section{Introduction}

Globular clusters lose stars mainly because of the dynamical processes like tidal disruption and dynamical friction \citep{fall77, fall85}. Both these processes can be studied by analyzing the stellar distribution of the globular clusters. Tidal disruption causes tidal tails which are aligned along and against the cluster motion, (e.g. Pal 5, for which \citet{kuzma15} found 30 stars in the trailing zone of the cluster and only 17 in the leading part). These tidal tails consist of cluster stars. This effect has been observed \citep{leon2000, Odenkirchen2001, Belokurov2006, Grillmair2006, jordi2010} and studied by many authors in the past \citep{king62, tremaine75, chernoff86, capdol93, weinberg94, meylan97, gnedin97, vesperini97, combes99, lotz01, capdel05, balbinot18}. The dynamical friction is due to the field stars that are concentrated in trailing part of the cluster opposite to its velocity vector. These field stars try to drag the cluster behind, slowing it down. This is called the gravitational wake effect. Dynamical friction has been predicted in many studies \citep{chandrasekhar43, mulder83, white83, tremaine84, capdol05b, TR, arca14} but has not been observed yet. A detailed study of the dynamical processes was not possible due to; (1) absence of proper motions, (2) relatively small fields of view on the sky, (3) absence of spectroscopic measurements of the tidal tail stars, like radial velocities and metallicities, to confirm membership.

Recently, \citet{kunder18} and \citet{danterrl} reported the detection of extra-tidal RR Lyrae stars in the bulge globular clusters NGC 6441 and NGC 6266 (M62), respectively. In this paper, we have analyzed the spatial distribution of extra-tidal RR Lyrae stars present in 56 Galactic globular clusters in order to study the effect of the dynamical processes on the clusters.

After the Gaia Data Release 2 \citep[Gaia DR2]{gaiadr2}, we have access to large amount of very accurate proper motion and magnitude (for G$<$19 magnitude) data. \citet{dr2acc} reported that Gaia DR2 proper motions deviate from Hipparcos \citep{lee07} and Tycho-2 \citep{tych02} proper motions mainly in the case of acceleration solutions and hence are very accurate for our purpose. This gives us an opportunity to explore the Solar neighborhood as well as probe deep into the space. In this analysis, we have used the Gaia data \citep{gaiarrl} to get proper motions of RR Lyrae stars along with globular cluster data from \citet[hereafter MPV14]{TR} and \citet[hereafter V18]{GC}. 

This paper is organized as follow: in section 2 we discuss the data used in the analysis. In section 3 we discuss the criteria of membership assignment. In section 4 we discuss the results. In section 5 we discuss the conclusions. 

\section{Data}

The data for our analysis have been taken from various sources. \citet{gaiarrl} published a Gaia DR2 catalog for RR Lyrae stars containing 140784 stars. Proper motion values for these 140784 RR Lyrae stars were extracted from Gaia DR2 using the Astronomical Data Query Language (ADQL) query \protect\footnote{\protect\url{ https://gea.esac.esa.int/archive/}}. V18 recently published a catalog containing the most accurate proper motions and distances for 150 globular clusters present in the Milky way, which were determined using Gaia DR2 data. Tidal radii values for 63 globular clusters are available from MPV14. Only 56 globular clusters were found to be common between MPV14 and V18. List of these common clusters is provided in Table~\ref{tab:t1}. Only these 56 common Galactic globular clusters were considered for further analysis.

\begin{table*}
	\centering
	\caption{Names of the clusters (ID), coordinates (RA and DEC), RA and DEC proper motions ($\mu$) and their respective errors ($\epsilon_{\mu}$) are from V18. Tidal radii (TR) and distances (Dist) for the clusters are from MPV14.}
	\label{tab:t1}
	\begin{tabular}{cccccccccc} 
\hline
 ID  &  RA  &  DEC  &  TR & Dist & $\mu_{RA}$ &  $\mu_{DEC}$ & $\epsilon_{\mu_{RA}}$ & $\epsilon_{\mu_{DEC}}$ & $\epsilon_{\mu_{RA}\mu_{DEC}}$ \\
   &  ($^{\circ}$)  &  ($^{\circ}$)  & (pc) &  (Kpc) & (mas/yr) & (mas/yr)  & (mas/yr)  & (mas/yr)  & (mas/yr)  \\
\hline
NGC 104 & 6.02363 & -72.08128 & 55.37 & 4.5 & 5.261 & -2.517 & 0.002 & 0.002 & -0.027 \\
NGC 362 & 15.80942 & -70.84878 & 26.61 & 8.6 & 6.721 & -2.519 & 0.007 & 0.005 & -0.04 \\
NGC 1904 & 81.04621 & -24.52472 & 30.9 & 12.9 & 2.475 & -1.574 & 0.006 & 0.007 & -0.033 \\
NGC 2808 & 138.01292 & -64.8635 & 25.35 & 9.6 & 1.018 & 0.27 & 0.006 & 0.006 & -0.076 \\
NGC 3201 & 154.40342 & -46.41247 & 36.13 & 4.9 & 8.336 & -1.982 & 0.003 & 0.003 & 0.09 \\
NGC 4147 & 182.52625 & 18.54264 & 34.16 & 19.3 & -1.702 & -2.108 & 0.016 & 0.01 & -0.147 \\
NGC 4372 & 186.43917 & -72.659 & 58.91 & 5.8 & -6.384 & 3.355 & 0.004 & 0.004 & 0.02 \\
NGC 4590 & 189.86658 & -26.74406 & 44.67 & 10.3 & -2.768 & 1.79 & 0.006 & 0.005 & -0.317 \\
NGC 4833 & 194.89133 & -70.8765 & 34.14 & 6.6 & -8.356 & -0.961 & 0.005 & 0.005 & 0.103 \\
NGC 5024 & 198.23021 & 18.16817 & 95.64 & 17.9 & -0.151 & -1.35 & 0.006 & 0.004 & -0.261 \\
NGC 5139 & 201.69683 & -47.47958 & 73.19 & 5.2 & -3.215 & -6.761 & 0.003 & 0.003 & -0.119 \\
NGC 5466 & 211.36371 & 28.53444 & 72.98 & 16.0 & -5.405 & -0.8 & 0.005 & 0.005 & 0.099 \\
Pal 5 & 229.02188 & -0.11161 & 51.17 & 23.2 & -2.728 & -2.687 & 0.022 & 0.025 & -0.39 \\
NGC 5897 & 229.35208 & -21.01028 & 36.88 & 12.5 & -5.408 & -3.457 & 0.007 & 0.006 & -0.143 \\
NGC 5904 & 229.63842 & 2.08103 & 51.55 & 7.5 & 4.08 & -9.867 & 0.004 & 0.004 & -0.037 \\
NGC 5927 & 232.00288 & -50.67303 & 37.45 & 7.7 & -5.064 & -3.207 & 0.01 & 0.009 & -0.11 \\
NGC 5986 & 236.5125 & -37.78642 & 24.15 & 10.4 & -4.186 & -4.573 & 0.01 & 0.008 & -0.154 \\
NGC 6093 & 244.26004 & -22.97608 & 20.88 & 10.0 & -2.941 & -5.588 & 0.012 & 0.01 & 0.008 \\
NGC 6121 & 245.89675 & -26.52575 & 33.16 & 2.2 & -12.484 & -18.983 & 0.005 & 0.004 & 0.023 \\
NGC 6144 & 246.80775 & -26.0235 & 86.35 & 8.9 & -1.736 & -2.631 & 0.009 & 0.007 & 0.085 \\
NGC 6171 & 248.13275 & -13.05378 & 35.33 & 6.4 & -1.909 & -5.976 & 0.008 & 0.006 & -0.04 \\
NGC 6205 & 250.42183 & 36.45986 & 43.39 & 7.1 & -3.172 & -2.586 & 0.004 & 0.005 & 0.152 \\
NGC 6218 & 251.80908 & -1.94853 & 24.13 & 4.8 & -0.149 & -6.776 & 0.005 & 0.004 & 0.27 \\
NGC 6254 & 254.28771 & -4.10031 & 23.64 & 4.4 & -4.728 & -6.551 & 0.005 & 0.004 & 0.142 \\
NGC 6266 & 255.30333 & -30.11372 & 23.1 & 6.8 & -5.058 & -3.02 & 0.013 & 0.011 & 0.109 \\
NGC 6273 & 255.6575 & -26.26797 & 37.3 & 8.8 & -3.241 & 1.671 & 0.009 & 0.007 & 0.152 \\
NGC 6284 & 256.11879 & -24.76486 & 102.72 & 15.3 & -3.192 & -2.035 & 0.015 & 0.013 & 0.103 \\
NGC 6287 & 256.28804 & -22.70836 & 19.02 & 9.4 & -4.98 & -1.886 & 0.011 & 0.008 & 0.066 \\
NGC 6293 & 257.5425 & -26.58208 & 39.32 & 9.5 & 0.882 & -4.314 & 0.014 & 0.012 & 0.106 \\
NGC 6304 & 258.63438 & -29.46203 & 22.74 & 5.9 & -4.039 & -1.043 & 0.017 & 0.014 & 0.16 \\
NGC 6316 & 259.15542 & -28.14011 & 22.97 & 10.4 & -4.942 & -4.598 & 0.021 & 0.018 & 0.117 \\
NGC 6333 & 259.79692 & -18.51594 & 18.39 & 7.9 & -2.214 & -3.216 & 0.007 & 0.006 & 0.226 \\
NGC 6341 & 259.28079 & 43.13594 & 30.05 & 8.3 & -4.923 & -0.556 & 0.005 & 0.006 & 0.085 \\
NGC 6342 & 260.292 & -19.58742 & 36.74 & 8.5 & -2.93 & -7.075 & 0.015 & 0.013 & 0.101 \\
NGC 6356 & 260.89554 & -17.81303 & 41.01 & 15.1 & -3.795 & -3.388 & 0.01 & 0.009 & 0.236 \\
NGC 6362 & 262.97913 & -67.04833 & 30.73 & 7.6 & -5.506 & -4.747 & 0.003 & 0.004 & 0.009 \\
NGC 6388 & 264.07179 & -44.7355 & 19.43 & 9.9 & -1.319 & -2.698 & 0.01 & 0.009 & 0.021 \\
NGC 6397 & 265.17538 & -53.67433 & 29.79 & 2.3 & 3.301 & -17.61 & 0.004 & 0.004 & -0.031 \\
NGC 6441 & 267.55442 & -37.05144 & 24.11 & 11.6 & -2.516 & -5.35 & 0.012 & 0.011 & 0.055 \\
NGC 6522 & 270.89175 & -30.03397 & 36.82 & 7.7 & 2.666 & -6.404 & 0.02 & 0.019 & 0.022 \\
NGC 6584 & 274.65667 & -52.21578 & 30.13 & 13.5 & -0.037 & -7.213 & 0.008 & 0.007 & -0.074 \\
NGC 6626 & 276.13671 & -24.86978 & 17.96 & 5.5 & -0.293 & -8.969 & 0.012 & 0.012 & -0.087 \\
NGC 6656 & 279.09975 & -23.90475 & 29.7 & 3.2 & 9.845 & -5.584 & 0.005 & 0.005 & 0.097 \\
NGC 6712 & 283.26792 & -8.70611 & 17.12 & 6.9 & 3.329 & -4.395 & 0.009 & 0.009 & 0.098 \\
NGC 6723 & 284.88813 & -36.63225 & 30.2 & 8.7 & 1.019 & -2.442 & 0.006 & 0.005 & 0.111 \\
NGC 6752 & 287.71712 & -59.98456 & 40.48 & 4.0 & -3.179 & -4.037 & 0.003 & 0.003 & 0.076 \\
NGC 6779 & 289.14821 & 30.18347 & 28.86 & 9.4 & -2.001 & 1.664 & 0.007 & 0.008 & 0.042 \\
NGC 6809 & 294.99879 & -30.96475 & 24.07 & 5.4 & -3.404 & -9.265 & 0.004 & 0.004 & 0.174 \\
NGC 6838 & 298.44371 & 18.77919 & 10.35 & 4.0 & -3.398 & -2.614 & 0.005 & 0.005 & 0.07 \\
NGC 6934 & 308.54738 & 7.40447 & 33.83 & 15.6 & -2.645 & -4.658 & 0.008 & 0.007 & 0.131 \\
NGC 7078 & 322.49304 & 12.167 & 63.23 & 10.4 & -0.622 & -3.782 & 0.006 & 0.006 & -0.037 \\
NGC 7099 & 325.09217 & -23.17986 & 37.43 & 8.1 & -0.705 & -7.215 & 0.007 & 0.007 & 0.269 \\
Pal 12 & 326.66183 & -21.25261 & 105.56 & 19.0 & -3.248 & -3.29 & 0.023 & 0.021 & 0.293 \\
NGC 1851 & 78.52817 & -40.04655 & 22.95 & 12.1 & 2.138 & -0.628 & 0.007 & 0.007 & -0.101 \\
NGC 5272 & 205.54842 & 28.37728 & 85.22 & 10.2 & -0.094 & -2.626 & 0.004 & 0.003 & -0.064 \\
Pal 13 & 346.68517 & 12.772 & 16.59 & 26.0 & 1.625 & 0.114 & 0.076 & 0.059 & 0.026\\
\hline
\end{tabular}
\end{table*}

\section{Method}

Starting with the Gaia DR2 catalog for RR Lyrae stars \citep{gaiarrl}, we selected only those stars for our analysis which seem to be the members of the clusters. To get these possible members, we selected RR Lyrae stars based on 3 criteria: \\
1. position of RR Lyrae stars in the sky close to the cluster;\\
2. proper motion consistent with the proper motion of the cluster; and \\
3. location in the CMD consistent with the CMD of the cluster.\\
Only those RR Lyrae stars which satisfy all the above criteria were selected as the members of the clusters. A description of implementing these criteria is provided below.

\subsection{Selecting extra-tidal RR Lyrae stars based on their position in sky}

To get the RR Lyrae stars which fulfill the first criteria, we defined an annulus around the center of the cluster with its lower radius as 2/3 times the tidal radius (smaller circle in Fig.~\ref{fig:f4}) and its upper radius as 3 times the tidal radius (bigger circle) in Fig.~\ref{fig:f4}) of the cluster. The RR Lyrae stars lying within the area bound by this annulus were selected as the extra-tidal RR Lyrae stars which are possible members of the cluster. Although the stars lying less than 2/3 times tidal radius away from the cluster center are also members of the cluster but for our analysis we are only interested in detecting the potential extra-tidal RR Lyrae stars which are present in the tidal tails of the cluster. Therefore, those RR Lyrae stars which are less than 2/3 times the tidal radius away from the center of the cluster are not taken into account. We have kept the criterion 1 uniform for all the clusters independent of their position in the sky because even if at this step some of the non-member stars are incorrectly selected due to the crowded field, they would be rejected in a further analysis.

\subsection{Selecting extra-tidal RR Lyrae stars based on their proper motions}

\begin{figure}
\begin{center}
\includegraphics[width=0.5\textwidth,keepaspectratio]{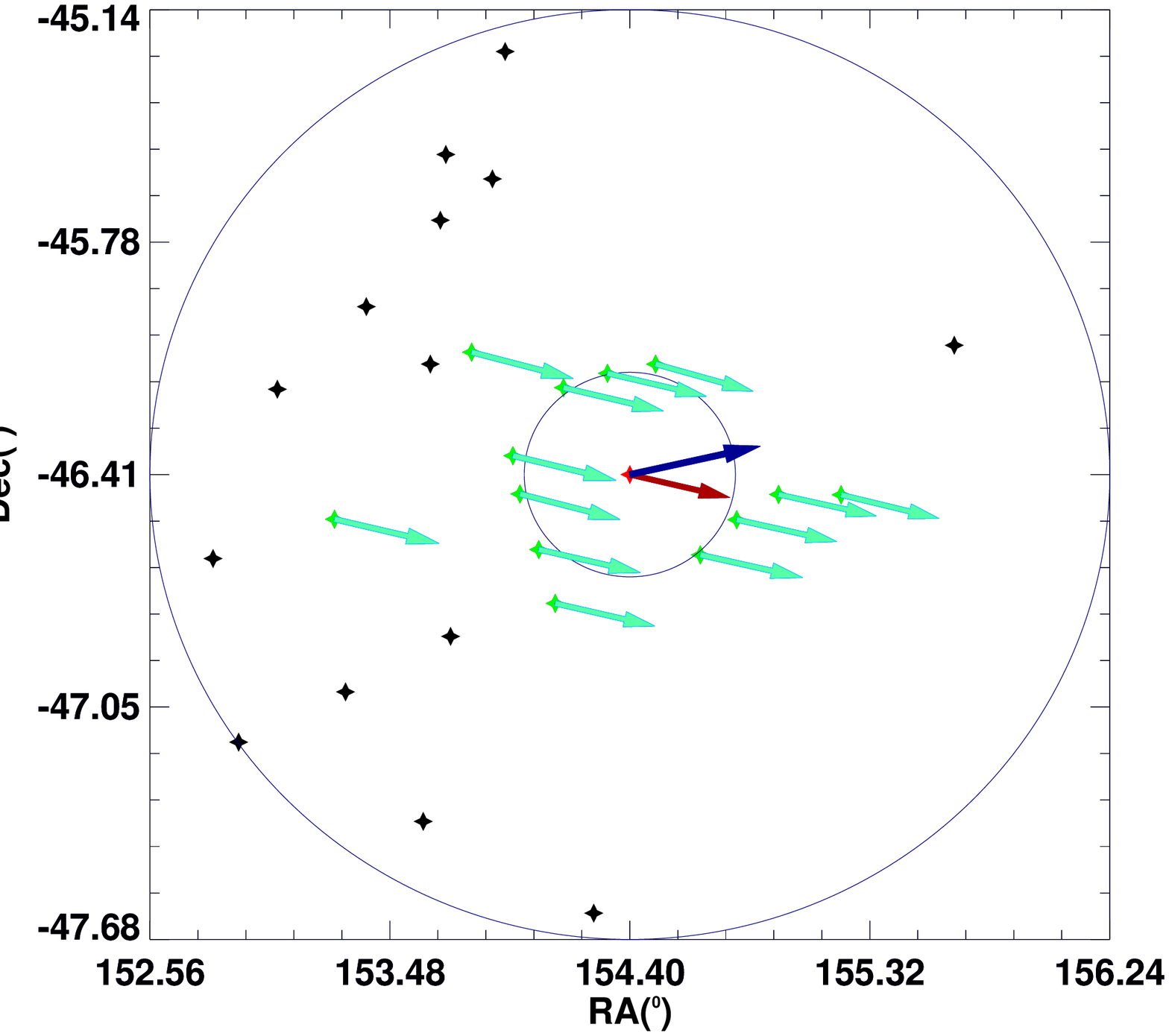}
\caption{Spatial distribution of the extra-tidal RR Lyrae stars present in the cluster NGC 3201 is shown. The RR Lyrae stars are selected (green dots) or rejected (black dots) as member of the cluster based on their proper motion. Proper motion of the cluster is indicated with maroon arrow. The smaller circle indicates the region which is $2/3$ times the tidal radius away from the cluster center and the bigger circle indicates the region which is $3$ times the tidal radius away from the cluster center. The proper motions of each of the selected RR Lyrae star (cyan arrows) are also shown. The blue arrow points towards the Galactic center.}
\label{fig:f4}
\end{center}
\end{figure}

\begin{figure}
\begin{center}
\includegraphics[width=0.5\textwidth,keepaspectratio]{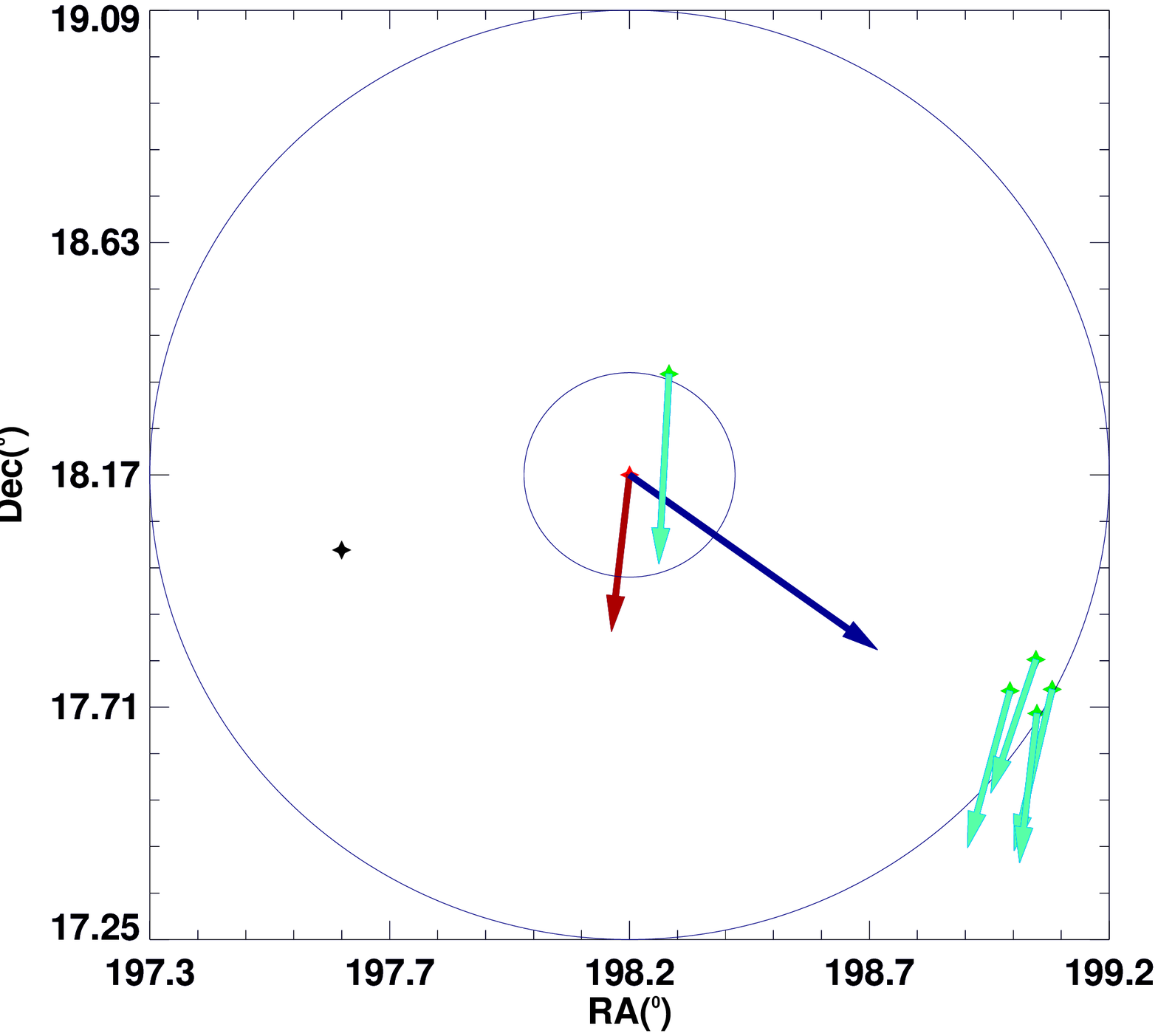}
\caption{Spatial distribution of the extra-tidal RR Lyrae stars present in the cluster NGC 5024. All the symbols and color representations are the same as in Fig.~\ref{fig:f4}.}
\label{fig:f7}
\end{center}
\end{figure}

If the cluster is in a crowded field then the non-member stars can be easily miss-classified as cluster members in the 2D space. Hence, the next criterion to select the RR Lyrae stars belonging to the cluster was based on the proper motion of the cluster. Therefore, once we have selected the RR Lyrae stars that are within $2/3$ to $3$ times the tidal radius away from the center of globular cluster, we next selected the stars which have approximately the same proper motion as that of the cluster. The proper motion of each selected RR Lyrae star was taken from Gaia DR2, and the mean proper motions for the clusters were taken from V18. We selected the stars for which the RA and DEC proper motions of the cluster match with the RA and DEC proper motions of the stars within 10 times the error in proper motion reported by Gaia. We allowed such large variation in the proper motion so that none of the stars which may be a member of the cluster is missed due to small error values. Errors in the proper motions of the clusters were negligible in comparison to the individual stellar proper motion errors. Therefore, these errors are neglected in the analysis. Selected stars from two clusters, NGC 3201 and NGC 5024 are shown in Fig.~\ref{fig:f4} and Fig.~\ref{fig:f7}, respectively. 

\subsection{Selecting extra-tidal RR Lyrae stars based on the CMD of the cluster}

\begin{figure}
\begin{center}
\includegraphics[width=0.5\textwidth,keepaspectratio]{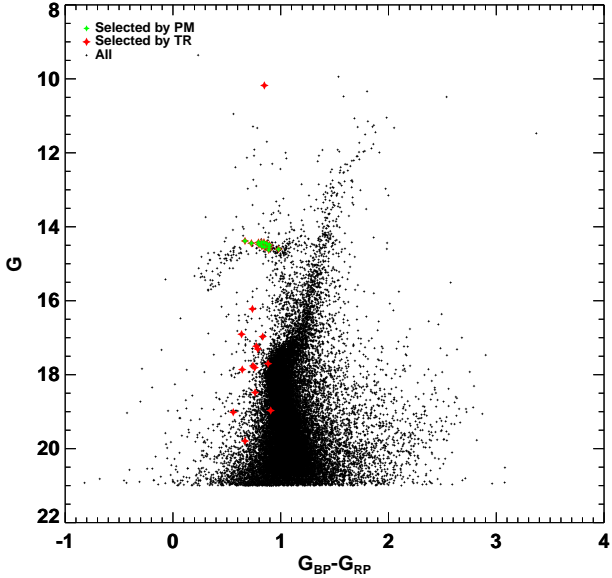}
\caption{Gaia DR2 CMD of the Galactic globular cluster NGC 3201. The green dots indicate the extra-tidal RR Lyrae stars selected on the basis of the first two criteria. The red dots indicate those RR Lyrae stars which were selected on the basis of their position in the sky but got rejected on the basis of their proper motions.}
\label{fig:f6}
\end{center}
\end{figure}

\begin{figure}
\begin{center}
\includegraphics[width=0.5\textwidth,keepaspectratio]{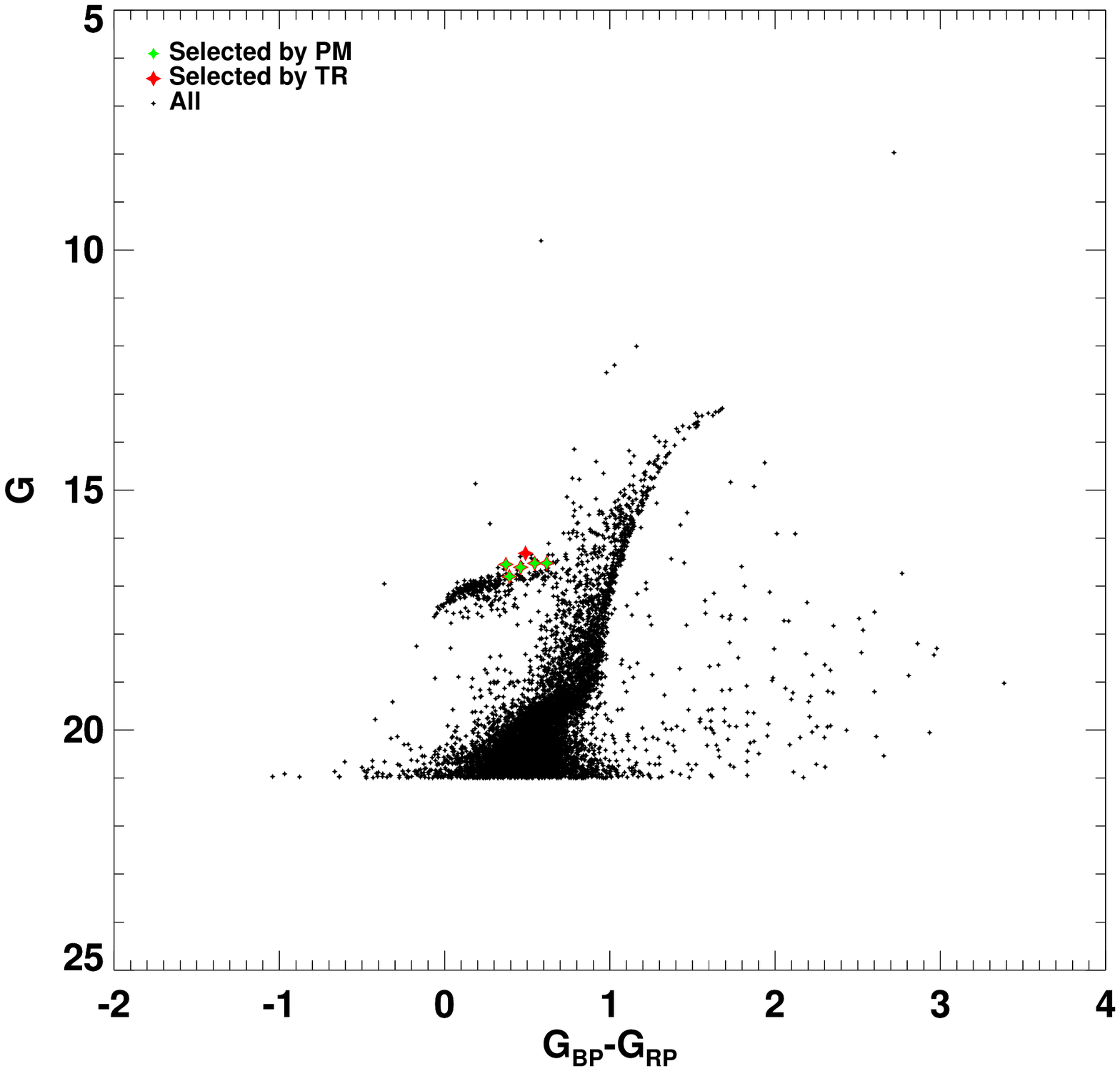}
\caption{Gaia DR2 CMD of the Galactic globular cluster NGC 5024. All the symbols and color representations are the same as in Fig.~\ref{fig:f6}.}
\label{fig:f8}
\end{center}
\end{figure}

While selecting extra-tidal RR Lyrae stars based on their position in the sky, we may have selected some of the stars that are part of some foreground and background source. In the proper motion analysis as well, we have allowed a variation of 10 times the error in the proper motion of each RR Lyrae star. Therefore, even after applying the first 2 criteria, we still may have selected some of the stars that are not cluster members. To be sure that we select only those RR Lyrae stars which are part of the cluster and not some background or foreground sources, we over-plotted our selected RR Lyrae stars on the CMD of the cluster. We selected only those RR Lyrae stars that lie on the intersection of the instability strip and the horizontal branch of the cluster CMD. Fig.~\ref{fig:f6} shows the CMD of NGC 3201. It can be seen from the figure that all the RR Lyrae stars that were selected on the basis of the first two criteria lie on the horizontal branch of the CMD of the cluster. Hence, these stars are part of the cluster. Similarly, the CMD of NGC 5024 (Fig.~\ref{fig:f8}) shows that all the RR Lyrae stars selected on the basis of the first two criteria lie on the horizontal branch of the CMD along with one star that was rejected based on its proper motion. Therefore, it is necessary to follow all the three criteria to make sure that all the selected RR Lyrae stars actually belong to the cluster. 

Instead of the CMD we could have used the Period-Luminosity (PL) relation for the RR Lyrae stars to select the members of the cluster. However, the  extinction corrections, the individual parallaxes and the Galactic PL relations are not accurate enough for our purposes. The location of the extra-tidal RR Lyrae stars on the CMDs is consistent with the CMD of the cluster, taking into account that some clusters are more reddened, and that differential reddening can also be present. Also, while the cluster CMDs are very well defined, some of the Gaia photometry in the crowded GC inner regions can have larger errors than the extra tidal stars that lie in the outskirts, or less crowded GC regions \citet{pancino17}.

A similar analysis was done for all the clusters. Table~\ref{tab:t3} lists the number of extra-tidal RR Lyrae stars selected at each step. The final results are listed in Table~\ref{tab:t2}. This table gives the parameters for the extra-tidal cluster RR Lyrae stars discovered here.

\begin{table*}

	\centering
	\caption{Number of extra-tidal RR Lyrae stars selected based on 3 criteria.}
	\label{tab:t3}
	\begin{tabular}{lc} 
\hline
Criteria & Number of stars selected\\
\hline
1. Stars around the center of the cluster within 2/3 and 3 times the tidal radius & 1531\\
2. Proper motion of the star and the cluster & 580\\
3. Position of the star on CMD of the cluster & 41\\

\hline
\end{tabular}
\end{table*}

\begin{table*}
	\centering
	\caption{Results of this work. Coordinates (RA and DEC), RA and DEC proper motions ($\mu$) and their respective errors ( $\epsilon_{\mu}$) taken from Gaia DR2 for each extra-tidal RR Lyrae star is provided along with its Gaia source ID and parent cluster.}
	\label{tab:t2}
	\begin{tabular}{cccccccc} 
\hline
Cluster & Gaia ID  &  RA  &  DEC  &  $\mu_{RA}$ & $\epsilon_{\mu_{RA}}$ &  $\mu_{DEC}$  & $\epsilon_{\mu_{DEC}}$ \\
   & &  ($^{\circ}$)  &  ($^{\circ}$)  & (mas/yr) & (mas/yr)  & (mas/yr)  & (mas/yr)  \\
\hline
NGC 1904   &  2957990469782163200  &       80.85  &      -24.46  &        2.49  &        0.05  &       -1.59  &        0.08  \\
NGC 3201   &  5413890427909525760  &      155.21  &      -46.47  &        8.14  &        0.04  &       -2.04  &        0.04  \\
NGC 3201   &  5413521335606162176  &      154.67  &      -46.63  &        8.49  &        0.04  &       -1.96  &        0.04  \\
NGC 3201   &  5413541539133152512  &      154.12  &      -46.76  &        8.23  &        0.04  &       -1.95  &        0.04  \\
NGC 3201   &  5413546349496406784  &      154.05  &      -46.62  &        8.43  &        0.04  &       -1.99  &        0.04  \\
NGC 3201   &  5413593731576291968  &      154.15  &      -46.18  &        8.36  &        0.03  &       -2.01  &        0.03  \\
NGC 3201   &  5413565178632931968  &      153.98  &      -46.47  &        8.32  &        0.03  &       -2.19  &        0.03  \\
NGC 3201   &  5413586133767131520  &      154.50  &      -46.11  &        8.16  &        0.03  &       -2.33  &        0.04  \\
NGC 3201   &  5413518926121264768  &      154.97  &      -46.47  &        8.14  &        0.03  &       -1.86  &        0.03  \\
NGC 3201   &  5413658224799006720  &      153.27  &      -46.53  &        8.68  &        0.04  &       -2.09  &        0.04  \\
NGC 3201   &  5413517487315507584  &      154.81  &      -46.54  &        8.32  &        0.03  &       -1.89  &        0.04  \\
NGC 3201   &  5413588096579420800  &      154.32  &      -46.14  &        8.28  &        0.03  &       -1.97  &        0.03  \\
NGC 3201   &  5413567927412022272  &      153.96  &      -46.36  &        8.59  &        0.03  &       -2.12  &        0.03  \\
NGC 3201   &  5413785424549983104  &      153.80  &      -46.08  &        8.50  &        0.04  &       -2.26  &        0.04  \\
NGC 4147   &  3950104443156396800  &      182.59  &       18.58  &       -1.66  &        0.17  &       -2.08  &        0.12  \\
NGC 5024   &  3938682995540541440  &      199.00  &       17.74  &       -0.35  &        0.15  &       -1.35  &        0.09  \\
NGC 5024   &  3938682445784738688  &      199.08  &       17.74  &       -0.32  &        0.16  &       -1.39  &        0.11  \\
NGC 5024   &  3939527386110983424  &      198.31  &       18.37  &       -0.08  &        0.17  &       -1.64  &        0.12  \\
NGC 5024   &  3938681827309432448  &      199.05  &       17.70  &       -0.14  &        0.15  &       -1.29  &        0.11  \\
NGC 5024   &  3938686637672823424  &      199.05  &       17.80  &       -0.38  &        0.14  &       -1.15  &        0.09  \\
NGC 5272   &  1454988929652528512  &      205.20  &       28.56  &       -0.21  &        0.07  &       -2.54  &        0.04  \\
NGC 5272   &  1454717693882616704  &      205.59  &       27.90  &       -0.41  &        0.06  &       -2.72  &        0.04  \\
NGC 5466   &  1452585014981202688  &      211.64  &       28.51  &       -5.61  &        0.11  &       -0.84  &        0.11  \\
NGC 6584   &  6701955277501739520  &      274.58  &      -52.31  &       -0.16  &        0.12  &       -6.99  &        0.11  \\
NGC 6584   &  6701962664852282368  &      274.80  &      -52.10  &       -0.09  &        0.11  &       -7.40  &        0.10  \\
NGC 6333   &  4122454132098289152  &      260.17  &      -18.65  &       -2.42  &        0.13  &       -3.07  &        0.09  \\
NGC 6341   &  1360194603384384256  &      259.37  &       42.75  &       -6.78  &        0.06  &       -0.70  &        0.05  \\
NGC 5139   &  6083892441183959168  &      202.43  &      -47.28  &       -4.42  &        0.09  &       -6.94  &        0.05  \\
Pal 5   &  4418734165978521728  &      229.14  &       -0.07  &       -2.58  &        0.25  &       -2.50  &        0.27  \\
Pal 5   &  4418913218870688768  &      228.99  &       -0.19  &       -2.76  &        0.23  &       -2.57  &        0.21  \\

\hline
\end{tabular}
\end{table*}

\section{Results}

Out of the 56 Galactic globular clusters analyzed here, only 11 have extra-tidal RR Lyrae stars that match all the three criteria. These 11 clusters have a total of 30 extra-tidal RR Lyrae stars. Most of these clusters have a couple of extra-tidal RR Lyrae stars. However, two of the clusters, NGC 3201 and NGC 5024 (also known as M 53), have enough extra-tidal RR Lyrae stars to draw some interesting conclusions. NGC 3201 (Fig.~\ref{fig:f4}) contains 13 extra-tidal RR Lyrae stars aligned asymmetrically along and against its direction of proper motion. Four of these RR Lyrae stars are located in the leading part of the cluster motion, while 7 of them are located in the trailing part. Based on these small number statistics, it appears that the tidal tails of this cluster are asymmetric. In the case of NGC 5024, the extra-tidal RR Lyrae stars are mostly concentrated in a small region which is at a distance of 3 times the tidal radius from the cluster center(Fig.~\ref{fig:f7}). The clusters with only a couple of extra-tidal RR Lyrae stars also have a high probability of having tidal tails, however, spectroscopic measurements, such as radial velocity and abundance information would give more insight into the nature and origin of the extra-tidal RR Lyrae stars..

NGC 3201 has an asymmetrical distribution of the extra-tidal RR Lyrae stars. This asymmetrical distribution of the extra-tidal RR Lyrae stars can be a combined effect of both dynamical friction and tidal disruption. However, to have a detectable effect of dynamical friction on the cluster, the cluster must be located in a region with high density of field stars. This kind of high density is only present in the bulge of the Galaxy and the cluster NGC 3201 is located well outside the bulge. Hence the most favourable argument to explain these asymmetrical tidal tails detected in the cluster NGC 3201 is the effect of tidal disruption combined with the stripped debris from the cluster. \citet{grillmair95} also detected extra-tidal extension for NGC 3201 based on optical surface density maps. \citet{chen10} analyzed the near-infrared stellar density profile of this cluster and found that this cluster shows stellar debris structure. They concluded that this stellar debris structure is due to the recent encounter of the cluster with the Galactic disk. \citet{kunder2014} also suspected the existence of tidal tails for NGC 3201, as they found some unbound stars located a few arc minutes outside of the cluster radius. Recently, \citet{anguiano16} reported the presence of tidal debris for the cluster. Their results were based on spectroscopic measurements from RAVE data. 

NGC 5024 has five extra-tidal RR Lyrae stars mostly concentrated in a region which is at a distance of 3 times the tidal radius from the center of the cluster. The presence of these extra-tidal RR Lyrae stars may be an indication of more extra-tidal population which may also belong to the cluster. \citet{beccari2010} found that in the case of NGC 5024, the model with lower concentration ($c=1.45$) and large core radius ($r_c=26"$) best reproduced the observed surface density profile in the inner portion of the cluster but this model deviates from observations in the outer region of the cluster. This type of deviation is a signature of the stars that are tidally stripping away from the cluster \citep{combes99, Johnston99,leon2000}.

\citet{kunder18} reported 8 extra-tidal RR Lyrae stars belonging to the Galactic globular cluster NGC 6441 but we did not find any in this study. This is due to three reasons: (1) some of the stars reported by them were not observed in Gaia DR2 (OGLE BLG-RRLYR-3774, OGLE BLG-RRLYR-30603, OGLE BLG-RRLYR-30792, OGLE BLG-RRLYR-3168, OGLE BLG-RRLYR-31080), (2) star reported was not marked as variable by Gaia survey (OGLE BLG-RRLYR-30513 (Gaia DR2 4039957151898674048)) and (3) stars were reported as RRab variable stars by Gaia survey but were not present in Gaia DR2 RR Lyrae catalog (catalog: gaiadr2.vari\_rrlyrae; stars: OGLE BLG-RRLYR-3613 (Gaia DR2 4039958564974057984)), OGLE BLG-RRLYR-30928 (Gaia DR2 4039978695412672768)).  \citet{danterrl} recently reported the presence of tidal tails in the cluster NGC 6266. These were not found in our analysis due to the fact that they used additional near-IR VVVX survey observations along with Gaia DR2 data.

\section{Conclusions}

We searched for extra-tidal RR Lyrae stars in 56 Galactic globular clusters based on the available literature data. Gaia DR2 data for RR Lyrae stars from the catalog of \citet{gaiarrl} was used along with data from MPV14 and V18 to assign cluster membership to each RR Lyrae star. The extra-tidal RR Lyrae stars were selected based on 3 criteria: position of the cluster and the star in the sky; proper motion of the cluster and the star; and position of the RR Lyrae star on the CMD of the cluster. All the three criteria are necessary and none of them can be neglected while selecting the extra-tidal RR Lyrae stars. Based on these three criteria we found 30 extra-tidal RR Lyrae stars belonging to 11 globular clusters. The majority of these clusters have only a couple of candidates. We also note that some of those extra stars for halo GCs may also be part of an extended halo \citep{kuzma16, kuzma18}. However, NGC 3201 and NGC 5024 have 13 and 5 extra-tidal RR Lyrae stars, respectively. It was found that in the case of NGC 3201, extra-tidal RR Lyrae stars are asymmetrically distributed around the cluster. We conclude that these asymmetrical tidal tails are due to the combination of tidal disruption with the tidal debris from the cluster. The tidal tails for this cluster have been indicated in the literature. However, in the case of NGC 5024 the extra-tidal RR Lyrae stars are mostly concentrated in a small region which is at a distance of 3 times the tidal radius away from the cluster center. The presence of extra-tidal stars in these clusters is an evidence that they are highly affected by tidal disruption and are disintegrating with time. 

The RR Lyrae stars are a small percentage of the numerous globular cluster population, such as main sequence stars. Therefore, the extra-tidal RR Lyrae stars detected in the clusters can be regarded as the tracers of more numerous population, which may have been members of the clusters in past and were ripped apart due to various dynamical processes. These stars could be revealed by deep wide field CMDs of the clusters. The detection of such stars would be helpful in understanding the cluster dynamics in detail.

\section*{Acknowledgements}

The authors thank the reviewer for a thorough review and highly appreciate the comments and suggestions, which significantly contributed to improving the content. RK is thankful to the Council of Scientific and Industrial Research, New Delhi, for a Senior Research Fellowship (SRF). RK would also like to thank Miss. Alessia Garofalo for helping her with the Gaia data during VOSS 2018. D.M.  gratefully acknowledges support provided by the BASAL  Center for Astrophysics and Associated Technologies (CATA) through grant PFB-06, and the Ministry for the Economy, Development and Tourism, Programa Iniciativa Cient\'ifica Milenio grant IC120009, awarded to the Millennium Institute of Astrophysics (MAS), and from project Fondecyt No. 1170121. HPS and RK thank the Indo-US Science and Technology Forum for funding the Indo-US virtual joint network centre on "Theoretical analyses of variable star light curves in the era of large surveys". RK and DM are also very grateful for the hospitality of the Vatican Observatory, where this work was started.




\bibliographystyle{rrl}








\bsp	
\label{lastpage}
\end{document}